\documentclass[referee,traditabstract]{aa}
\usepackage{amssymb}
\usepackage{graphics}
\usepackage{tabularx}
\usepackage{color}
\usepackage{natbib}
\usepackage{rotating}

\newcommand{\quasar}{3C454.3}

\begin{document}

\title{Search for signatures of dust in the Pluto--Charon system using 
\emph{Herschel}\thanks{Herschel is an ESA space observatory with science instruments 
provided by European-led Principal Investigator consortia and with important participation 
from NASA.}/PACS observations}

\author{G. Marton\inst{1} \and Cs. Kiss\inst{1} \and Z. Balog\inst{2} \and E. Lellouch\inst{3} \and E. Vereb\'elyi\inst{1} \and U. Klaas\inst{2}}

\institute{%
Konkoly Observatory, Research Centre for Astronomy and Earth Sciences, 
Hungarian Academy of Sciences, Konkoly Thege Mikl\'os \'ut 15-17, 1121 Budapest, Hungary
\and
Max-Planck-Institut f\"ur Astronomie, K\"onigstuhl 17, D-69117 Heidelberg, Germany
\and Observatoire de Paris, CNRS, UPMC Univ. Paris 06, Univ. Paris-Diderot, 5
 place Jules Janssen, 92195 Meudon Principal Cedex, France}

\abstract{In this letter we explore the environment of Pluto and Charon in the far infrared with the main aim to identify the signs of any possible dust ring, should it exist in the system. Our study is based on observations performed at 70$\mu$m with the PACS instrument onboard the Herschel Space Observatory at 9 epochs between March 14 and 19, 2012. The far-infrared images of the Pluto-Charon system are compared to those of the point spread function (PSF) reference quasar \quasar. The deviation between the observed Pluto--Charon and reference PSFs are less then 1$\sigma$ indicating that clear evidence for an extended dust ring around the system was not found. Our method is capable of detecting a hypothetical ring with a total flux of $\sim$3.3 mJy at a distance of $\sim$153 000 km ($\sim$8.2 Pluto--Charon distances) from the system barycentre. We place upper limits on the total disk mass and on the column density in a reasonable disk configuration and analyse the hazard during the flyby of NASAÕs New Horizons in July 2015. This realistic model configuration predicts a column density of 8.7$\times10^{-10}$ g cm$^{-2}$ along the path of the probe and an impactor mass of 8.7$\times10^{-5}$ g.}

\authorrunning{G. Marton et al.}
\titlerunning{Search for signatures of dust in the Pluto--Charon system}
\maketitle
\sloppy

\section{Introduction}

The Pluto--Charon system harbours at least four smaller satellites. Nix and Hydra 
\citep{weaver2006, buie2006} have diameters estimated to be around 100 km or less, 
while Kerberos and Styx \citep{showalter2011, showalter2012} have diameters less than 
40\,km. Studies of the phase space \citep{nagy2006, suli2009, giuliatti2010, 
zsigmond2010, pires2011, giuliatti2013} showed that dynamically stable regions exist around the 
system. Formation of a debris disk as a result of collision of small bodies was also 
suggested \citep{durda2000, steffl2006}. The existence of such a ring was 
investigated by \citet{steffl2007} but observational evidence for the ring was not found. They placed a 3$\sigma$  upper limit on the azimuthally averaged normal I/F of ring particles of 5.1$\times10^7$ at a distance of 42\,000 km from the Pluto-Charon barycentre. \citet{boissel2014} derived an upper limit of 15\,000 for the number of bodies smaller than 0.3 km at distances smaller than 70\,000 km from Pluto's system barycentre. The impactor velocity 
of colliding bodies ranges from 10--100\,m\,s$^{-1}$ \citep{steffl2007}. Given that 
the escape velocities of Nix and Hydra are between 30 and 90\,m\,s$^{-1}$, 
any material ejected from these bodies can escape from the 
satellites but cannot escape from the system (Pluto and Charon escape velocities are $\sim$1200 and $\sim$500\,m\,s$^{-1}$, respectively). Recently a ring system around the centaurs Chariklo \citep{braga2014} was discovered and several indicators show that probably (2060) Chiron \citep{ortiz2015} harbours a ring system, too. These suggest that dust rings around small bodies may be a common feature.

The existence of a dust ring around the system would also mean a hazard for NASA's 
'New Horizons' probe \citep{stern2008}. The goal of the mission is to provide the first in situ
 exploration of the system, with a close fly-by in July 2015. 

Visual wavelength range reflected light observations \citep{showalter2011, showalter2012} taken with the 
Hubble Space Telescope (HST) are, however, not ideal for the detection 
of diffuse material in the Pluto--Charon system due to the very high 
brightness of Pluto and Charon and a possibly low surface brightness of the hypothetical 
ring. In the thermal infrared the 
contrast between the central object (star) and the surrounding material (debris disk)
is higher than in the visual range, due to the very different spectral energy distributions (SEDs) of the two 
components. In the case of Pluto, the thermal infrared may still be beneficial in the case of dark grains in the hypothetical ring, like e.g. the ones found in the Uranus system \citep[albedo of $\sim$6\%,][]{karkoschka2001}. 

In this letter we investigate thermal emission (70\,$\mu$m) images of the Pluto--Charon system, taken with the Photodetector Array Camera and Spectrometer \citep[PACS,][]{poglitsch2010} of the Herschel Space Observatory \citep[][]{pilbratt2010}, in order to identify any extended emission, originating from sources other than Pluto and Charon themselves. 

Diffuse dust structures like disks and rings modify the PSF of their parent sources, 
even if they cannot be fully resolved. Such partially resolved features -- if bright enough -- 
cause a broadening of the PSF of the source. This kind of broadening has been used 
to detect partially resolved debris disks with the PACS photometer in many cases 
\citep[see e.g.][]{moor2015}. In the case of PACS, the 70$\mu$m band has the highest spatial resolution (FWHM of 5\farcs7), and
also the SED of large grains peaks around this wavelength at the heliocentric distance of Pluto. 
In the case of smaller grains the temperature is higher than Pluto's, 
causing a shortward shift of the peak of the SED, again favouring the shortest wavelength PACS band
to detect it. The ratio of the fluxes from a disk/ring and Pluto is expected to be the highest 
in this band as well. Therefore we only use the 70$\mu$m band data in the following analysis.

\section{Observations and data reduction}

The PACS camera on board the Herschel observed the 
Pluto--Charon system multiple times. We used the light-curve measurement of 
Lellouch et al. (proposal ID: OT2\_elellouc\_2, \citet{lellouch2}  \& Lellouch et al. 2015, in prep.) 
taken at 70/160\,$\mu$m between March 14 and 19 in 2012. All measurements 
were made in scan-map mode with a scan speed of 20$^{\prime\prime}\,s^{-1}$ and 
lasted for 286\,s. Pluto--Charon was observed at a heliocentric distance of $r$\,=\,32.2\,au, 
and at a distance of $\Delta\simeq$32.4\,au from the telescope. 
Basic characteristics of the Pluto observations are listed in Table ~\ref{obstable}.

In addition to the target measurements, we used the engineering observations of the fiducial star 
$\alpha$\,Boo for calibration purposes, as well as 42 Herschel/PACS 
measurements of the quasar \quasar{} \citep[proposal ID: TOO\_awehrle\_2, see ][]{wehrle2012} as
point spread function reference (discussed in details in Sect.~\ref{plutomodeling}). 
The quasar observations were taken between November 19, 2010 and January 10, 2011. 
Each observation was 220\,s long, in the same observational 
mode and with the same scan-speed as in the case of the Pluto measurements.

All observations were processed with the Herschel Interactive Processing Environment 
\citep[HIPE,][]{ott2010}, version 13 (developer release). Starting from the Level-1 
stage the observations were first corrected for the evaporator temperature effect 
\citep[see][]{moor2014}, with correction factors of $<$0.3\% to 3.2\%. 
Then we followed the standard pipeline steps to eliminate the 1/f noise as described in \citet{balog2014} using iterative masked high-pass filtering of the 
observational timelines with the {\it photProject()} task. For large scale extended 
emission studies the mapmaker algorithm JScanam is often more suitable, but in our case 
the structures are kept in the inner $\sim$30\arcsec of the source, so there is no need to use JScanam to deal with a broadened PSF. For more details see the mapmaker report of \citep{paladini2013}. 
We mosaic the images with 1.1$^{\prime\prime}$ pixel sizes. As Pluto was
moving relatively fast w.r.t. the sky reference frame ($\sim$2.1\arcsec\,h$^{-1}$),
we constructed the final images in Pluto's co-moving reference frame for each individual 
OBSID. We also used temperature correction in order to eliminate effects caused by 
the telescope mirror temperature variations. We eliminated the slight pointing jitter by using 
the experimental tool that corrects the pointing based on the information provided by 
the gyroscopes of the satellite. To alleviate the effect of correlated noise we used 
the drizzle parameter of {\it pixfrac}\,=\,0.2 
\citep[see][for more details on the PACS photometer noise characterisation]{popesso}.

\section{Thermal emission model of the Pluto--Charon system \label{plutomodeling}}

The observed PSF of any object is a superposition of monochromatic PSFs over the wavelength range of the bandpass filter. The shape of the object's SED determines the observed PSF FWHM. Objects with high temperature have a decreasing SED in the PACS 70 $\mu$m band and monochromatic PSFs of the shorter wavelengths are dominant. A constant or rising SED at these wavelengths obviously leads to the dominance of longer wavelength PSFs, resulting in a broadened FWHM value with respect to e.g. that of a normal star. 

In the thermal infrared the SED of Pluto is, indeed, very different from that of a stellar photosphere, due to the very different temperatures. Calibration stars of the PACS photometers have temperatures of 3000\,K or higher, while the surface temperature of Pluto and Charon are 
$\sim$50\,K \citep{Lellouch}. Even the bright minor planet Vesta, whose PSF was widely used in different Herschel applications \citep{lutz2012}, has a typical surface temperature of $\sim$200\,K, notably higher than that of Pluto. The spectral index $\alpha$ of Pluto is $\sim$0.82, assuming a mean $T_b$ of 46.5K and 43K at 70 and 160 $\mu$m \citep{lellouch2}, where $\alpha(\nu)\,=\,\frac{\partial\,log\,S(\nu)}{\partial\,log\,\nu}$. For comparison, the spectral indices of a stellar photosphere and that of Vesta are $\sim$2 at this wavelength.

\begin{figure}
\includegraphics[width=8.3cm,trim=1cm 0.5cm 1.5cm 1cm]{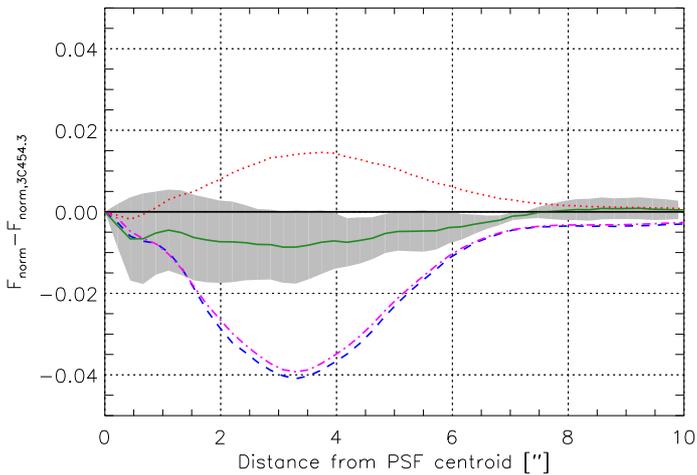}
\caption{Normalised intensity of \quasar{} subtracted from the normalised intensity of the object PSFs, as a function of distance from the 
PSF centroid. The reference profile is the PSF of quasar 3C454.3 (black solid [0,0] line). 
The calibrator star profile (blue dashed line) is the narrowest. The Vesta PSF (dashed magenta) is wider than the calibrator star, but still narrower than the reference PSF. The filled grey area represents the uncertainty of the 9 Pluto observations. The mean median (MM) Pluto profile is shown as the green solid line. The most realistic dust ring model 
detected at 1+3$\sigma$ significance level is plotted as the red dotted line. \label{psfdiff}}
\end{figure}

To exclude the possibility of misinterpreting the SED-broadened Pluto PSF as a signature of a ring, we have collected the observations of the quasar \quasar{}\footnote{http://herschel.esac.esa.int/hpt/publicationdetailsview.do?\newline bibcode=2012ApJ...758...72W} which is known to have a spectral index of 0.88 $\pm$ 0.07 at 70 -- 160 $\mu$m \citep{wehrle2012}, very similar to that of Pluto. The reference PSF was created by combining the 42 observations of this target. The cosmological distance of \quasar{} ensures that the region where its thermal emission originates
from is small enough that it cannot be resolved by PACS at 70\,$\mu$m, hence can be used as a low temperature point source reference PSF. 
Using this combined PSF a high-resolution PSF template with 0\farcs22 pixel size was created. This template is used in the following steps to 
construct the synthetic PSFs (see below). The combined quasar PSF has been compared with the combined PSF of the
fiducial star and we found that the fiducial star PSF is obviously narrower (see Fig.~\ref{psfdiff} and Fig.~\ref{psfdist}). 

For each epoch a synthetic Pluto PSF was constructed using the combined quasar PSF as template. 
The synthetic Pluto PSF has two components, one corresponding to Pluto itself,
the other describing the contribution of Charon (see Eq.~\ref{eq:flux}). 
The relative positions of these two bodies were taken from the Horizons database, 
as listed in Table~\ref{obstable} and as presented in Fig.~\ref{relpos}. In the high spatial resolution 
model images the two sources are placed in these relative positions. 
The relative contribution of Pluto and Charon to the total flux 
are characterised by the $X$ ratio:

\begin{eqnarray}\nonumber
F(x,y) = XF_{q}(x,y)+(1-X)F_{q}(x-x_0,y-y_0)
\label{eq:flux}
\end{eqnarray}

\noindent where F$_q$ is the normalised synthetic quasar PSF and x$_0$ and 
y$_0$ are the offsets of Charon's position w.r.t. Pluto. The original, high 
resolution image is convolved with the synthetic quasar PSF to obtain
an "observable" model PSF. Due to the rotational variability 
of the surface of Pluto and Charon, the relative strength of the thermal fluxes 
cannot be determined very accurately, but their ratio is close to
2:1 or 3:1 depending on the model of their thermal properties \citep{Lellouch}. In our investigation we used both ratios, i.e X=0.67 and X=0.75, respectively.
The radial intensity profiles of the synthetic Pluto+Charon PSFs agree
with each other very well, indicating that the position and relative contribution
of Charon to the total flux of the system has a minor effect on the
intensity profile. This is mainly driven by the fact that Charon is very close
to Pluto (0\farcs6--0\farcs8) compared to the size of the PSF in this band (5\farcs7).
In this respect, the difference observed between the 
radial intensity curves of the individual Pluto observations should mostly 
be attributed to measurement uncertainties. Therefore we constructed a mean
measured (MM) radial intensity curve of the 9 Pluto measurements, using 
equal weights in the case of all measurements (note that the integration times and 
other conditions were the same for all measurements). 

The agreement of the model and observed curves are characterised by the
reduced $\chi^2$-value:
\begin{equation}
\chi^2 _r= \frac{1}{N-N_{DOF}-1}\sum\limits_i{{(y_i^j - y_i^m)^2}\over{\delta_i^2}}
\label{eq:s}
\end{equation}
\noindent where N is the number of data points, N$_{DOF}$ is the degrees of freedom, $y_i^j$-s are the normalised intensity 
of the j-th model at $i$ distance and $y_i^m$ are the median observed normalised intensities. 
The $\delta_i$ values are the standard deviations of the observed normalised
intensities at a specific radial distance. 

\section{Intensity profile results}\label{ipr}

When the MM curve of Pluto is compared with the mean synthetic (MS) curve of the 9 epochs,
we obtain a difference with a low level of significance (0.8$\sigma$) between radial distances 0\arcsec{}--22\arcsec{}. This shows that the observed Pluto 
PSFs are very similar to the synthetic ones and that \emph{we are not able to
detect any extended emission feature around Pluto, i.e. there is no obvious 
evidence for the existence of a notable dust ring in the Pluto--Charon system.}

\begin{figure}
\includegraphics[width=8.5cm,trim=1cm 0.5cm 1cm 1cm]{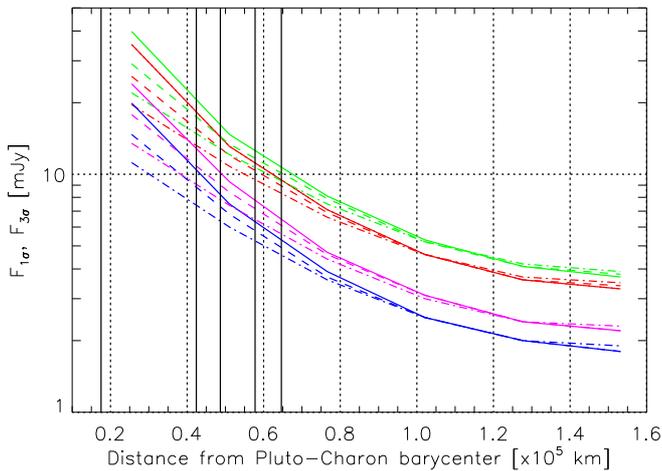}
\caption[]{F$_0$ (ring excess) values as a function of distance from the Pluto--Charon barycentre.
Red and green colours correspond to F$_{Pluto}$ and F$_{Charon}$ ratio of 
2:1, and 3:1, for 3$\sigma$ significance. Blue and magenta colours refer to the same ratios, but for 1$\sigma$.  
Solid, dashed and dashed-dotted lines represent the width of the modelled ring 
${\Delta}r$\,=\,0\farcs5, 1\farcs0 and 1\farcs5, respectively. Vertical solid black lines from left to right represent the orbital semi-major axis of Pluto moons Charon, Styx, Nix, Kerberos and Hydra.}

\label{fig:ring_excess}
\end{figure}

It is, however, an important question, what the total brightness 
of such a hypothetical disk/ring could be that would still be detectable with our 
method. To test this, we constructed a set of model images, in each case 
placing a ring in addition to the contributions of Pluto and Charon, 
as described above. The moons of the Pluto system seem to revolve in a well-defined 
plane, hence any stable ring/disk is expected in this plane with the
highest probability \citep{nagy2006}. We model the ring as a surface brightness enhancement around a specific distance ($r_0$) from the barycentre of the Pluto-Charon system in the midplane of the moon system (tilted w.r.t. the line of sight by an orientation angle of 61$^{\circ}$), assuming a Gaussian intensity profile in the surface brightness distribution. 
The parameters characterising the surface brightness of the 
ring are (i) r$_0$, the distance of the ring from the barycentre, (ii)
the width of the Gaussian $\Delta r$ and (iii) F$_r$, the total flux density of the
ring. We constructed model images using these parameters in the
range F$_r$\,=\,0--60\,mJy, $r_0$\,=\,2.55$\times10^4$--1.53$\times10^5$ km (1\farcs1--6\farcs6) and $\Delta r$\,=\,1.16$\times10^4$, 1.74$\times10^4$ and 2.32$\times10^4$ km (0\farcs5, 
0\farcs75 and 1\farcs0). For each model we calculate the $\chi^2_r$ of the difference. 
The F$_0$ (ring excess) dependence of the $\chi^2_r$ values corresponding to the $1\sigma$ and $3\sigma$ significance levels are presented in Fig.\ref{fig:ring_excess}, where $\sigma$ is the standard deviation of the $\chi^2$ distribution. 

Fig. \ref{fig:ring_excess} shows that our method is the most sensitive at higher radial distances, and the detection probability also depends on the X flux ratio. In the case of rings close to Pluto the detection limit is significantly higher. E.g. at the distance of Charon, we would only be able to detect a ring with $\sim$100\,mJy excess, therefore we are not able to constrain the presence / lack of a ring within or close to the orbit of Charon. The lowest detection limit belongs to X\,=\,0.67 with ring distance of 6\farcs6, in this case F($1\sigma$)\,=\,1.8\,mJy and F($3\sigma$)\,=\,3.3\,mJy.

A possible dust ring around the Pluto--Charon system is expected to be in the region where the satellites exist. We assume that the most realistic model is where X=0.67, r$_0$\,=\,2\farcs2 (5.1$\times10^4$ km) and $\Delta r$\,=\,1\farcs0 (2.32$\times10^4$ km). In this case the ring is centered at a distance between Kerberos and Nix and the ring $\Delta r$ also models an extended feature of which 58\% exists inside the orbit of the Pluto satellites. The results show that for this ring configuration, we obtain F$_0$\,=\,6.0 and 10.8\,mJy for the $1\sigma$ and $3\sigma$ significance levels,
respectively. This implies that the detection probability for this geometry is $\sim$68\% 
for a ring with 6.0\,mJy total flux density and 99.7\% with 10.8\,mJy. 


\section{Discussion}
Our method presented above is able to detect \emph{infrared excess} from a ring according to Fig. 2. Any calculation that wants to derive dust surface density or impactor mass based on this excess \emph{has to rely} on assumptions on the type, spatial and grain size distribution of the dust.

Assuming that the material of our hypothetical ring is dominated by grains which are in thermal equilibrium with the solar radiation, 
we can give a rough estimate of the mass of this ring. Considering a black body temperature of T$_d$\,=\,$278.3\times(r/1[au])^{-1/2}$\,=\,48.9K \citep{kennedy} and an absorption coefficient of 
$\kappa_{abs}$ =  5.96 m$^2$ kg$^{-1}$ at 70\,$\mu$m \citep{li2001} we can estimate the dust mass 
according to Eq.~\ref{eq3} below:

\begin{equation}
M_{dust}=\frac{\Delta^2 F_{\nu}(\lambda)}{\kappa_{abs} B_{\nu}(\lambda,T)} = 
2.28\times10^{9}\times\frac{F_{70}(\lambda)}{mJy} [kg],
\label{eq3}
\end{equation} 

\noindent where $\Delta$ is the distance to the observer and B$_{\nu}$(T$_d$) is the Planck function. 
Applying Eq. \ref{eq3} to the most realistic case put forward in Section \ref{ipr}, the 3$\sigma$ flux limit is 10.8\,mJy and we obtain a total disk mass of 2.5$\times10^{10}$ kg, equivalent to the mass of a $\sim$340\,m-sized small body assuming an average density of $\rho$\,=\,1.2\,g\,cm$^{-3}$.

The flyby of New Horizons is planned to take place inside the Charon orbit. 
Taking the maximum separation of 0\farcs85 between Pluto and Charon we calculated the fraction of mass and the corresponding surface density for the area between the two objects. The fraction of the total ring dust mass located within 0\farcs85 is 4.3$\times10^{-4}$ (1.1$\times10^{7}$\,kg). This corresponds to a surface density of 8.7$\times10^{-10}$ g\,cm$^{-2}$ and an
impactor mass of $\sim$8.7$\times$10$^{-5}$\,g. This
is very similar, only $\sim$10\,\% lower,  than the impactor mass of 10$^{-4}$\,g, obtained by \citet{steffl2007} based on Hubble Space Telescope observations. We note that a thinner ring with the same brightness would result in a significantly lower impactor mass.

\section{Summary}

In this paper we searched for far-infrared signatures of a possible dust ring around the 
Pluto--Charon system. 
We found no clear observational evidence for a dust ring but we were able to put model-dependent upper limits on the
possible amount of dust emission that could remain undetectable by our method. For a realistic ring configuration we put a $3\sigma$ upper flux density limit of 10.8\,mJy at 70\,$\mu$m. This corresponds to a surface density of 8.7$\times10^{-10}$ g\,cm$^{-2}$ in the region between Pluto and Charon. Our estimated impactor mass for the New Horizons flyby is very similar ($\sim10\%$ lower) to the previous results of \citet{steffl2007} and suggests that the satellite will not be endangered by particle collisions.


\begin{acknowledgements}
We thank our anonymous referee for the useful comments which helped to improve the content of the paper. This work has been supported by the PECS contract \#\,4000109997/13/NL/KML of the 
Hungarian Space Office and the European Space Agency; and the Hungarian Research Fund (OTKA) grants nrs. 101939 and 104607. 
\end{acknowledgements}

\appendix
\section{Basic properties of the Pluto--Charon Herschel observations}

\begin{figure*}
\includegraphics[width=8.8cm]{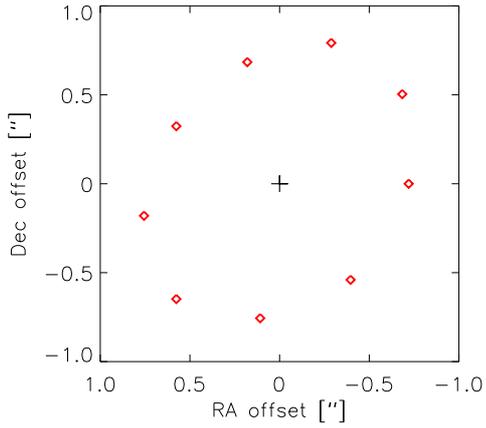}
\caption{Position of Charon (red rhombs) at the time of the 
observations relative to Pluto (black plus sign). \label{relpos}}
\end{figure*}

\begin{table*}
\small
\begin{tabular}{crrr}
 OBSID & Start Time & PA$_{PC}$ & Sep.    \\
       &           & [deg] &  [$^{\prime\prime}$]  \\ 
\hline
1342241381 & 2012-03-14 03:02:01 & 59.21 & 0.63\\
1342241382 & 2012-03-14 03:07:50 & 59.21 & 0.63\\
1342241418 & 2012-03-14 19:54:46 & 104.15 & 0.74\\
1342241419 & 2012-03-14 20:00:35 & 107.63 & 0.71\\
1342241471 & 2012-03-15 12:58:39 & 140.0 & 0.85\\
1342241472 & 2012-03-15 13:04:28 & 140.0 & 0.85\\
1342241509 & 2012-03-16 06:33:59 & 172.32 & 0.76\\
1342241510 & 2012-03-16 06:39:48 & 172.66 & 0.80\\
1342241620 & 2012-03-17 00:10:12 & -145.30 & 0.66\\
1342241621 & 2012-03-17 00:16:01 & -143.43 & 0.63\\
1342241655 & 2012-03-17 17:31:54 & -90.0 & 0.68\\
1342241656 & 2012-03-17 17:37:43 & -93.19 & 0.65\\
1342241699 & 2012-03-18 11:04:14 & -52.03 & 0.82\\
1342241700 & 2012-03-18 11:10:03 & -52.58 & 0.77\\
1342241865 & 2012-03-19 04:42:46 & -18.95 & 0.84\\
1342241866 & 2012-03-19 04:48:35 & -19.78& 0.80\\
1342241928 & 2012-03-19 20:44:31 & 13.95 & 0.70\\
1342241929 & 2012-03-19 20:50:20 & 11.24 & 0.70\\ 
\hline
\end{tabular}
\caption{Basic properties of the Pluto--Charon Herschel observations. 
All observations lasted for 286\,s (single repetition). 
The columns are: (1) Observation identifier; 
(2) Date and start time of the observation; 
(3) Position angle of Charon w.r.t. Pluto;
(4) Separation of Pluto and Charon;\label{obstable}}
\end{table*}

\section{Modelled total ring flux values}

\begin{table*}
\begin{tabular}{c|ccc|ccc}
\hline
\multicolumn{7}{c}{1$\sigma$}\\
\hline
Flux ratio & \multicolumn{3}{c|}{2:1} & \multicolumn{3}{c}{3:1} \\
\hline
$\Delta$r & 0\farcs5	& 0\farcs75	&  1\farcs0	&  0\farcs5	& 0\farcs75	&  1\farcs0\\
\hline
Distance & \multicolumn{3}{c|}{Flux [mJy]} & \multicolumn{3}{c}{Flux [mJy]} \\
1\farcs1	&19.9	&14.7	&11.2	&24.0	&17.8	&13.5\\ 
2\farcs2	&7.5		&6.8		&6.0		&9.3		&8.4		&7.4\\ 
3\farcs3	&3.9		&3.7		&3.6		&4.7		&4.6		&4.4\\ 
4\farcs4	&2.5		&2.5		&2.5		&3.1		&3.1		&3.0\\ 
5\farcs5	&2.0		&2.0		&2.0		&2.4		&2.4		&2.4\\ 
6\farcs6	&1.8		&1.8		&1.9		&2.2		&2.2		&2.3\\
\hline

\hline
\multicolumn{7}{c}{3$\sigma$}\\
\hline
Flux ratio & \multicolumn{3}{c|}{2:1} & \multicolumn{3}{c}{3:1} \\
\hline
$\Delta$r & 0\farcs5	& 0\farcs75	&  1\farcs0	&  0\farcs5	& 0\farcs75	&  1\farcs0\\
\hline
Distance & \multicolumn{3}{c|}{Flux [mJy]} & \multicolumn{3}{c}{Flux [mJy]} \\
1\farcs1	&35.1	&	25.8		&	19.6		&	39.8		&	29.1		&	22.0\\
2\farcs2	&13.1	&	12.0		&	10.8		&	14.7		&	13.4		&	12.1\\
3\farcs3	&7.1		&	6.9		&	6.6		&	8.1		&	7.9		&	7.5\\
4\farcs4	&4.6		&	4.6		&	4.6		&	5.3		&	5.2		&	5.2\\
5\farcs5	&3.6		&	3.6		&	3.7		&	4.1		&	4.1		&	4.2\\
6\farcs6	&3.3		&	3.4		&	3.5		&	3.7		&	3.8		&	3.9\\
\hline
\end{tabular}
\caption{Total flux values of model rings of different configurations. We placed the model dust ring with a Gaussian intensity profile at different distances from with varying FWHM and Pluto--Charon flux ratios.}
\label{fluxtable}
\end{table*}

\section{PSF profiles}
\begin{figure*}
\includegraphics[width=8.8cm]{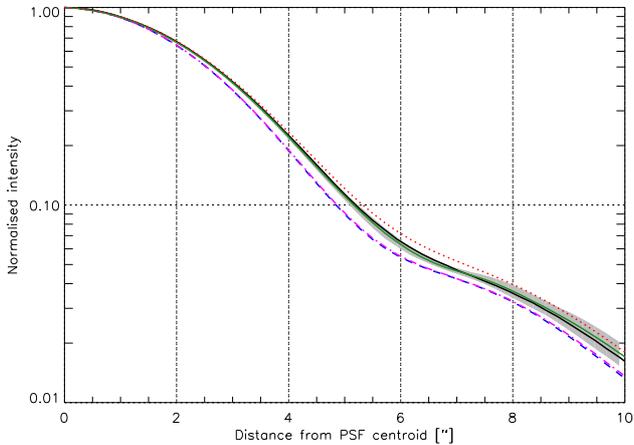}
\caption{Normalised intensity profiles as a function of distance from the 
PSF centroid. The quasar \quasar  \,is shown with black solid line. The calibrator star profile is presented with blue dashed line. The dashed-dotted magenta line indicates the Vesta PSF. The solid green line shows the mean of the 9 Pluto observations (MM) while the filled grey area represents the measurement uncertainty. The faintest dust ring model detected at 3$\sigma$ significance level is plotted with dotted red line. \label{psfdist}}
\end{figure*}

\end{document}